# IN SITU PLASMA PROCESSING OF SUPERCONDUCTING CAVITIES AT JEFFERSON LAB*

T. Powers, N. Brock, and T. Ganey
Thomas Jefferson National Accelerator Facility, Newport News, VA, USA

## Abstract

Jefferson Lab began a plasma processing program starting in the spring of 2019. Plasma processing is a common technique for removing hydrocarbons from surfaces, which increases the work function and reduces the secondary emission coefficient. Unlike helium processing which relies on ion bombardment of the field emitters, plasma processing uses free oxygen produced in the plasma to break down the hydrocarbons on the surface of the cavity. The residuals of the hydrocarbons in the form of water, carbon monoxide and carbon dioxide are removed from the cryomodule as part of the process gas flow. The initial focus of the effort is processing C100 cavities by injecting RF power into the HOM coupler ports. We will then start investigating processing of C50 cavities by introducing RF into the fundamental power coupler. The plan is to start processing cryomodules in the CEBAF tunnel in the mid-term future, with a goal of improving the operational gradients and the energy margin of the linacs. This work will describe the systems and methods used at JLAB for processing cavities using an argon oxygen gas mixture. Before and after plasma processing results will also be presented.

## BACKGROUND

The main accelerator at Jefferson Lab is the Continuous Electron Beam Accelerator Facility (CEBAF), which is a continuous wave 5-pass recirculating accelerator that uses 418 superconducting cavities to accelerate electrons to a maximum design energy of 12 GeV. The cavities are configured into 52-1/4 cryomodules. Twelve of the cryomodules are known as C100 style cryomodules which use TESLA-style HOM coupler antennas. The remaining 40-1/4 cryomodules are known as C20, C50 and C75 style cryomodules. These cryomodules have internal ceramic HOM loads, which are waveguide coupled to the beam line and are operated at 2K, with no RF connections outside of the cryomodule.

Many of the cavities in CEBAF suffer from field emission due to particulates on the surface of the cavities. While some of these particles originate from the production processes, which have improved substantially in the past 30 years, a fraction of the particles is assumed to be transported into the cavities via the beamline during normal operation and maintenance. These introduced particles can cause new field emitters to "turn on" which further degrades machine operation [1]. For the most recent C100 rebuild, which was installed in the fall of 2019, seven of the eight cavities had no field emission and the rebuilt module has been routinely operated at 100 MeV with no measurable radiation.

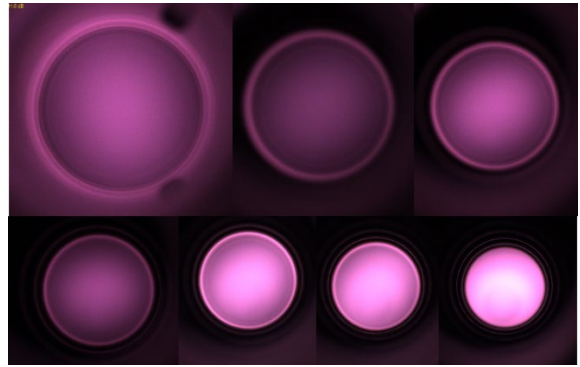

Figure 1: Images of argon plasma in each of the 7 cells of a C100 Cavity.

In the C100 cryomodules and most of the C50/C75 cryomodules, while field emission does not cause frequent arcing, it does cause extra heat load to the helium system. Unfortunately, due to physics needs for higher energy, the cavities are normally operated at relatively high levels of field emission. This causes significant radiation primarily at the ends of the cryomodule which, in addition to activating beam line hardware, damages components such as O-ring seals and instrumentation cables.

### Plasma Processing of SRF Cavities

Plasma processing using argon/oxygen gas mixtures is used in a number of industrial applications for removing hydrocarbons from the surfaces prior to the application of thin films or performing surface analysis [2]. Unfortunately, due to the size and shape of the SRF structures in a cryomodule, the gas pressure and the free oxygen recombination cross section, these industrial systems are not suitable for this application.

One of the early applications using a mixture of a noble gas and a small percentage of oxygen with positive results on was in 2012 at the Synchrotron Radiation Center located at University of Wisconsin, where the WiFEL SRF gun cathode surface fields were improved from 6 MV/m to 26 MV/m [3]. ORNL has an ongoing program of processing cryomodules in SNS using a neon/oxygen gas mixture. To date they have processed 36 cavities *in situ* with an average improvement of 2.5 MV/m [4, 5]. Staff at Fermilab started a program three years ago that is focused on processing LCLS II cavities via one of the HOM ports [6].

___________

* Funding provided by SC Nuclear Physics Program through DOE SC Lab funding announcement Lab-20-2310 and U.S. DOE, Office of Science, Office of Nuclear Physics under contract DE-AC05-06OR23177.
powers@jlab.org

## Current Jefferson Lab Program

A new plasma processing program was initiated at JLAB in the spring of 2019, with initial plasma production in a C100 cavity in August 2019, a series of laboratory bench top measurements was initiated in the fall of 2019. A series of vertical test, plasma process, vertical test cycles started in the Fall of 2020 and a cryomodule, which was removed from the tunnel for rebuild, was processed in the cryomodule test facility in June 2021. The plan is to continue the vertical test series as well as the off-line development through FY22 and to start processing cryomodules in the tunnel as soon as it is practical.

## METHODS

In this method a discharge is established, and electrons with an energy between 10 and a few hundred eV break the bond between oxygen atoms and free oxygen is produced. The free oxygen oxidizes the hydrocarbons on the surface of the cavities and the fragments leave the cavity together with the process gas, which is set up as a continuous flow. Fig. 1 shows images of plasma in each of the cells of a C100 cavity.

### Gas System

The initial plasma processing method that we are using at JLAB is similar to that used at ORNL. Process gas pressures between 80 and 200 mTorr are used, the mass flow is set up between 5 and 30 SCCM. The oxygen percentage is set to levels between 0.5% and 3%. The gas passes through a 10 nm cartridge filter just prior going into the clean portion of the gas manifold or to the cavity. The process pressure is measured using a 1 Torr full scale capacitance manometer.

The pumping cart contains two turbo molecular pumps, a primary pump and second pump, which is used to maintain the proper pressure on the RGA manifold. The pressure is regulated with an up-stream mass flow controller and the mass flow is adjusted with throttling valves upstream of the main turbo pump. Using this method, we can regulate the pressure to a few percent and maintain constant mass flow to about ±5%. The RGA pressure is set using a leak valve. In addition to monitoring the argon to oxygen ratio, the RGA is used to monitor $H_2$, CO, $CO_2$ and $H_2O$, which are hydrocarbon fragments that are produced when the free oxygen interacts with the hydrocarbons.

### RF Systems

Because the intrinsic $Q_0$ of a warm C100 cavity is $10^5$ and design Q of the fundamental power coupler (FPC) is $3 \times 10^7$, only about 4% of the power that is applied to the FPC makes it into the cavity. Fortunately, the HOM antennas on C100 cavities are designed to have a low external-Q at HOM frequencies. For this reason, we have chosen to inject RF power into the cavity using the first 5 TE111 HOM modes. Unlike the TM110 modes, which are dipole modes, these modes produce a uniform plasma in the center of the cells.

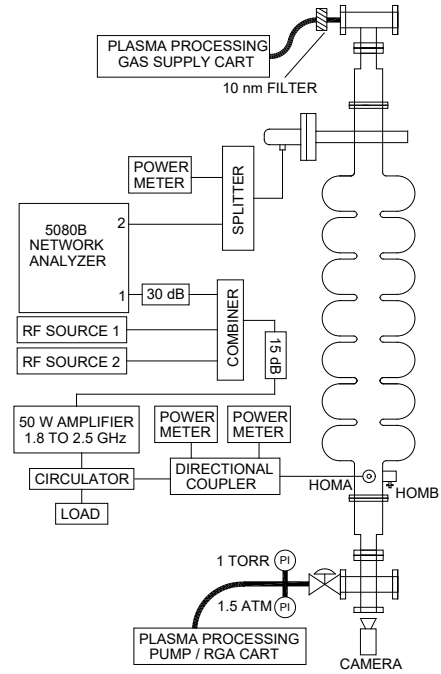

Figure 2: Setup for plasma processing in vertical test area.

A diagram of the RF system used for processing a C100 cavity is shown in Fig. 2. The signals from two RF sources are combined with output of a network analyzer which is set up to sweep through the TE111 and TM110 modes. The output of this combiner is applied to the input of a 50 W RF amplifier the output of which is applied to the cavity. With this network analyzer setup one is able to measure the frequency shifts caused by the plasma and use it as a metric of plasma density and location [7]. As part of this program and in preparation for processing multiple cryomodules in the CEBAF tunnel we built up two RF systems that are capable of processing two cavities at once and a third, R&D system, that is capable of processing one cavity at a time.

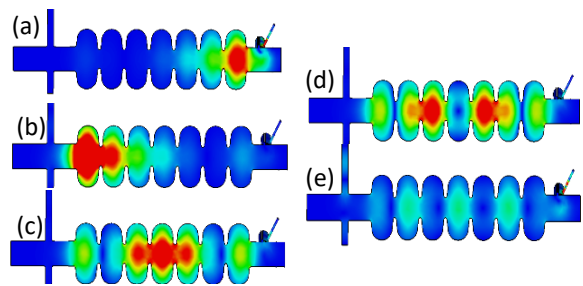

Figure 3: Results of simulations of the first 5 TE111 cavity modes [7].

### Choosing the Modes

At first, simple symmetrical gradient distribution patterns were used. This was followed up by a series of simulations that included the resistive nature of the surfaces and the end conditions of the model. The color mapped cavities in Fig. 3 are the results of the more realistic simulations using mixed boundary conditions.

In general, each mode has two polarizations anchored by the location of the HOM couplers. These couplers are separated by 90° and are rotated 45° as compared to the location of the fundamental power coupler. For the TM110 dipole modes this resulted in a non-uniform plasma with higher densities, which are rotated to the angles of the HOM couplers. Figure 4 is an image of such a mode. Figure 1 shows the more uniform TE111 Modes.

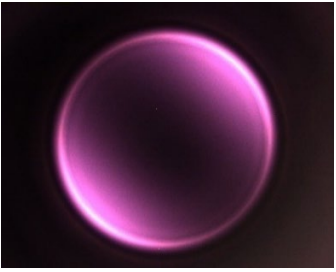

Figure 4: Image of a dipole mode in cell 4.

Figure 5: is the results of standard bead pull measurements with a dielectric bead where, in general, the relative gradient can be approximated as Eq. (1):

$$E(z) \propto \sqrt{tan\left(\Delta S_{21-phase}(z)\right)(1 - |S11|)} \qquad (1)$$

Where $E(z)$ is the relative gradient at location z, $|S11|$ is the magnitude of the S11 voltage ratio at the resonant frequency and, $\Delta S_{21-phase}(z)$ is the phase difference between S21 with the bead removed from the cavity and when the bead is at the location z. We used this data to determine that the desired polarization is one where there is a maximum S21, e.g. stronger electric field in the cell next to the fundamental power coupler, while still having a reasonable minimum of S11(dB).

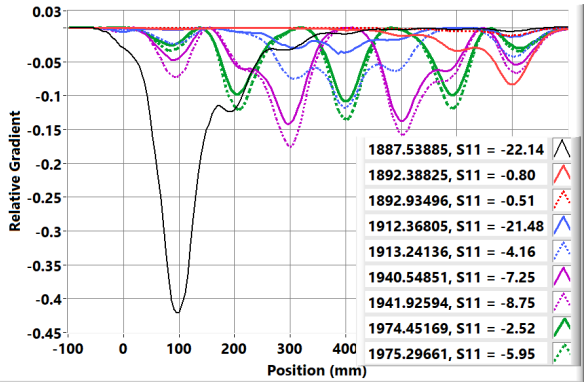

Figure 5: Relative gradient for the different polarizations of the first 5 TE111 modes.

## Moving From Cell to Cell

Our standard recipe for establishing the plasma is to use $2\pi/7$ mode frequency to get a field pattern shown in Fig. 3(c) to establish a plasma in cell 4, where cell 1 is adjacent to the HOM couplers. Once in cell 4, we could go to cell 7 by applying the $1\pi/7$ mode frequency for the pattern in Fig. 3(b). Then to cells 6 and 5 by applying the $4\pi/7$ mode frequency, Fig. 3(d), then $3\pi/7$ mode frequency, Fig. 3(e), in order. Alternately we could go from cell 4 to 3 to 2 to 1 by applying the frequencies for the modes shown in Figs. 3(d), 3(e), then 3(a). Except for the transition from cell 4 to cell 7, in all cases, because of Debye shielding which lowers the downstream EM fields, the plasma always moved towards the source of RF. In most cases we processed the cavities in pairs of 1/2, 3/4, 5/6, with cell 7 processed separately.

## HOM Coupler Breakdown

We experienced HOM coupler breakdown early on in the project while we were learning how to establish a plasma and how to understand which modes excited which cells. Breakdowns were easy to detect when we were using a camera but, initially, they were difficult to detect using just RF power meters.

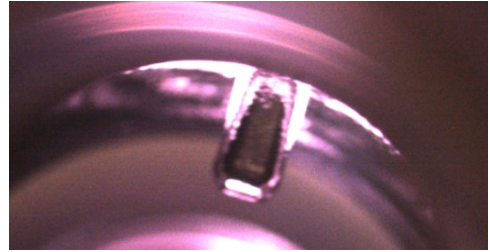

Figure 6: Image of a coupler breakdown.

Figure 6 is an image of a coupler with a local discharge. In this case the breakdown is in the region of the RF feedthrough antenna. The antenna that was used on that particular series of tests was observed to be pitted at the end of the tests. The assumption is that this breakdown was occurring because of the inner surface of the HOM feedthrough tube was very rough. The assumption is that this roughness was due to using BCP to process the cavity and that the acid was trapped in that small volume. Breakdown events were also observed on the HOM antenna tips. Both types of faults can be detected using RF by measuring the ratio of the incident power divided by the transmitted power as there as the transmitted power is substantially reduced during an HOM breakdown. Additionally, there is a significant change in the network analyzer trace when there is a coupler breakdown.

## Implications of Adjacent HOM Couplers

Unlike TESLA-style cavities which have an HOM coupler at each end of the cavity, in C100 cavities both HOM couplers are at the end opposite the fundamental power coupler and the tips of the antennas are approximately 2 cm apart. Because of this, there is a strong coupling between the two HOM ports especially in the frequency range of the TE111 modes. Additionally, when processing in a cryomodule there is a cable that is 244±2.5 cm long on each of the HOM couplers. Thus, any signal that is coupled from the driven HOM coupler to the passive HOM coupler travels down the cable and is reflected back with a fixed but random phase and has the potential to affect the fields within the structure.

All of the initial processing work at JLAB was done with an unterminated passive HOM coupler that did not have a long cable attached to it. Thus, for all of that work the reflected wave behaved relatively uniformly. When we did some initial experiments with a long cable we found that we could no longer ignite a plasma and we had frequent HOM coupler breakdowns.

We investigated two methods to address this issue. The first was to use an RF splitter, and drive each of the HOM ports with the same signal. With this, there is an uncertainty of the relative phase of the two HOM ports due to the tolerance of the length of the cables within the cryomodule. The second was to control the phase of the return signal by using a mechanical phase shifter on the open cable, which was connected to the other HOM coupler.

The first set of measurements involved calibrated S11 and S21 measurements of the system as a function of phase shift. In the case, where we drove both couplers, the phase shifter was placed between the splitter and one of the HOM ports. In the other case where we drove just one coupler, the phase shifter was placed at the end of an RF cable of the same length that is used in the cryomodule. The results for the case where we drove both couplers in combination with bead pull data indicated that a 5° difference in phase would cause a substantial change in the field patterns. Thus, when driving both couplers it would be difficult to determine the proper phase shifter setting such that all of the modes can be excited by the RF source without adjusting the phase shifter.

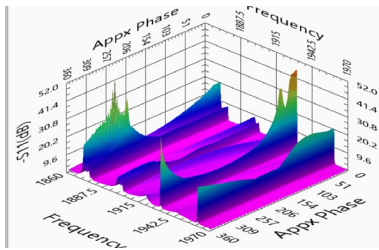

Figure 7: -S11 for phase shifter at the end of a 240 cm long cable.

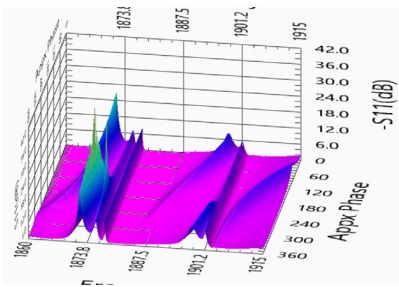

Figure 8. -S11 for the same cavity showing the first three modes, at as well as other "modes" which are likely due to the unterminated cable.

Figures 7 and 8 show -S11(dB) as a function of frequency and phase shift of an open phase shifter that is placed on the end of a 240 cm cable of the same type that is used in a cryomodule. In Fig. 8 the three straight "ridges" at 1875.6, 1877.8 and 1901.4 MHz are modes of the cavity. The small ridges with minimum values of S11 of 7 dB that change in magnitude as a function of frequency between 1915 and 1902 MHz as well as between 1901 and 1887 MHz are likely due to the two-way losses in the 240 cm cable. The "mode" that changes in frequency below 1873.8 MHz is not well understood but preliminary data indicates that it produces non-symmetrical fields in the cell closest to the HOM couplers.

We used S11 and S21 data taken in this manner to determine the phase shifter setting for processing cavities in the cryomodule. The goal was to use cavity modes that were well away from the modes that changed in frequency as a function of phase shifter position. For example, for the cavity in Fig. 8, a phase shift of 120 degrees was selected.

### Software

A suite of semi-automated software was developed and used for processing cavities in a cryomodule, vertically, and in the off-line test facility. In addition to vacuum controls the software was configured to record the information from an RGA, the RF systems, and a network analyzer which was used to measured S21 while processing a cavity. Automatic ignition software as well as determination of frequency shift patterns due to the plasma were also included in all three setups. For the vertical and off-line systems, the tools included image capture so that we could verify the excited cells as well as coupler breakdowns. The off-line system also included software for an optical spectrometer which was used to understand the discharge physics.

## RESULTS

Two types of results are reported here. The first is vertical test results where a cavity was tested vertically, plasma processed in the vertical staging area and the vertical test was repeated. In these tests the cavity remained on the vertical test stand. The second set of results is from processing a cryomodule. This cryomodule was C100-10 which had been in the accelerator for several years and which suffered from severe degradation due to particle movement within the beamline, most likely due to a damaged beam line valve.

### Vertical Processing and Test Results

To date, the vertical test program was done with one 7-cell C100 cavity. The initial purpose of the vertical test and processing program was to debug the system and validate procedures. In the future, the program will be used to develop novel techniques and processes. Additionally, since it is not possible to use a camera when processing a cryomodule a secondary purpose was to develop the processes necessary for processing a cryomodule.

Figure 9 shows the results of the second series of vertical tests. In this the red triangles are the results of the vertical tests prior to processing. The blue circles are plots of the results after two cycles of plasma processing. Field emission (FE) onset increased by 1 MV/m. The operating gradient with 100 mRem/hr of radiation was increased from 16.8 MV/m to 19.5 MV/m. The higher $Q_0$ values on the second test were attributed to the fact that this cavity was nitrogen doped about six years prior, that the first cooldown was not a rapid cooldown and the second cooldown was a fast cooldown.

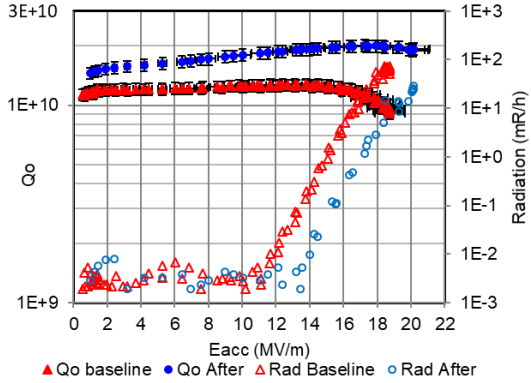

Figure 9: Results from the second series of vertical cavity processing cycles.

### Cryomodule Processing Results

One of the primary functions of processing a cryomodule in an off-line facility was to validate the equipment and software as well as to develop the processes and procedures necessary for processing a cryomodule. To that end, we chose a cryomodule that was removed from the tunnel for rebuild. It was chosen to be rebuilt because there were Viton O-ring failures on two beam line valves one about 2 m upstream of cavity 1, followed by one about 30 cm upstream of cavity 1. While the cryomodule operated well after the first failure, cavities 1 and 2 had to be bypassed after the second failure, which was not fully diagnosed until 18 months after latter. Thus, in addition to being a pro cess development opportunity, this can be considered an experiment in the effectiveness of plasma processing of a cryomodule under extreme conditions. We processed each cell or pair of cells for one hour, waited at least 48 hours, then processed the cavity again. The exception is cavity 1 which was processed about 8 hours per day for 5 days. Using two channels of RF we were able to process 8 cavities in two nine-hour shifts. Beamline vacuum was re-established within 3 hours after stopping the process flow and opening the main valves on the pumping cart.

Table 1: Gradients (MV/m) for CM100-10 before and after processing.

| Cav | E (MV/m) at FE onset | | E (MV/m) at 100 mR/hr | | E (MV/m) Max | |
|---|---|---|---|---|---|---|
| | Before | After | Before | After | Before | After |
| 1 | 4.3 | 3.6 | 4.7 | 4.1 | 6.5 | 5.3 |
| 2 | 5.1 | 5.3 | 5.7 | 5.9 | 8.0 | 7.8 |
| 3 | 6.3 | 6.8 | 6.9 | 7.0 | 10.9 | 11.5 |
| 4 | 5.1 | 5.3 | 6.3 | 5.7 | 7.8 | 8.1 |
| 5 | 6.6 | 8.2 | 7.2 | 9.0 | 11.5 | 15 |
| 6 | 6.5 | 6.4 | 7.1 | 6.8 | 9.3 | 9.9 |
| 7 | 7.2 | 9.1 | 7.8 | 10.1 | 9.6 | 15 |
| 8 | 5.6 | 5.0 | 6.2 | 5.5 | 6.9 | 8 |

Table 1 shows the before and after cryomodule results, where FE onset is the gradient where the onset was where the radiation measured at the boundary of the cryomodule exceeded 10 mR/hr. As somewhat expected, the improvement to the cryomodule after processing was minimal with a maximum improvement in field emission onset of 1.9 MV/m and two cavities showing a measurable decrease in field emission onset. Those cavities happened to be at the ends of the cryomodule. Overall there was an increase in the operational voltage of the cryomodule from 36.3 MV to 37.9 MV.

## SUMMARY

A status of the plasma processing program at Jefferson Lab has been presented. We obtained reasonable improvements in field emission onset up to 1.3 MV/m in our vertical test program where the cavities started out in reasonable condition. We have demonstrated the ability to process a cryomodule and that this is not a miracle cure for badly contaminated structures. The next step is to develop methods for processing C50 and C75 cryomodules as well as to start to investigate novel techniques and methods that have the potential to address contaminants other than hydrocarbons. Processing of cryomodules in the CEBAF accelerator is planned in the mid-term future.